\documentstyle[aps]{revtex}
\begin{document}
\draft

\def\ket#1{|#1\rangle}
\def\bra#1{\langle#1|}
\def\scal#1#2{\langle#1|#2\rangle}
\def\matr#1#2#3{\langle#1|#2|#3\rangle}

\title{Parameter symmetries of quantum many-body systems}
\author{
Pavel Cejnar$^{1,2}$\thanks{Electronic address: pavel.cejnar@mff.cuni.cz}
and
Hendrik B. Geyer$^1$\thanks{Electronic address: hbg@sunvax.sun.ac.za}
}
\address{
$^1$Institute of Theoretical Physics, University of Stellenbosch,\\
7602 Matieland, South Africa\\
$^2$Institute of Particle and Nuclear Physics, Charles University,\\
V Hole\v sovi\v ck\'ach 2, 180\,00 Prague, Czech Republic
}
\date{\today}
\maketitle

\begin{abstract}

We analyze the occurrence of dynamically equivalent Hamiltonians in
the parameter space of general many-body interactions for quantum
systems, particularly those that conserve the total number of
particles.  As an illustration of the general framework, the
appearance of parameter symmetries in the interacting boson model-1
and their absence in the Ginocchio SO$_8$ fermionic model are
discussed.

\end{abstract}

\section{Introduction}

It is generally accepted that symmetry belongs to the most
fundamental concepts in physics. In particular, the generalization
of the standard invariance groups in terms of dynamical (spectrum
generating) groups and dynamical symmetries \cite{Barut} seems to
provide a rather general framework for describing both classical
and quantum physical systems~\cite{Dothan}. The role of dynamical
symmetries in problems of quantum integrability is reviewed by Zhang
and Feng in Ref.\cite{Zhang}. One of the well known examples of the
algebraic approach in nonrelativistic quantum physics is the family of
so-called interacting boson models (IBM) , introduced by Arima and
Iachello \cite{Iachello} and extensively employed in phenomenological
nuclear physics.
The dynamical groups of these models are easily tractable and they
neatly decompose into separate dynamical-symmetry chains, each
having a clear geometric interpretation and an associated set of
nuclei conforming to the various symmetry dictated predictions.

There is, however, a certain ambiguity in the definition of some of
the IBM dynamical symmetries resulting from possible gauge
transformations of boson operators in the symmetry limits
\cite{Iachello,Isacker}.  This ambiguity applies even to the simplest
version of the model, the IBM-1, where the choice of the boson gauge
was for long considered as a mere convention.  It was, however,
recently recognized as a deeper and universal property of general
algebraic systems \cite{Kusnezov}.  Because the twin symmetries
resulting from the gauge transformation can be located \lq\lq
between\rq\rq\ standard symmetries in the parameter space, i.e.\
seemingly in transitional regions, they were referred to as \lq\lq
hidden\rq\rq\ \cite{Kusnezov}.  The consequences of these \lq\lq
hidden\rq\rq\ symmetries for the problem of quantum chaos were
emphasized in Ref.\cite{Cejnar}.

In the work by Shirokov {\it et al.\/} \cite{Shirokov1} gauge
transformations of boson operators, and associated hidden symmetries,
were studied from the more general perspective of what these authors
call \lq\lq parameter symmetries.\rq\rq\ It was shown that each IBM
Hamiltonian has an isospectral partner located at a different point in
parameter space.  Hidden symmetries emerge as special cases of these
parameter symmetries -- they arise when the parameter symmetry partner
is constructed for a Hamiltonian possessing a dynamical symmetry.
Subsequently the idea was also utilized within the two-component
proton-neutron interacting boson model, IBM-2 \cite{Shirokov2}.  It is
clear, however, that parameter symmetries can be explored in a much
wider class of parameter-dependent systems.  It is therefore the aim
of the present work to discuss the occurrence of parameter
symmetries in more general situations.

We first analyze some generic features of parameter symmetries
(Sec.~\ref{general}) and their realization in many-body systems which
conserve the total number of particles (Sec.~\ref{manybody}).  Two
concrete examples are then considered in detail, namely the
interacting boson model-1 (Sec.~\ref{ibm}) and the Ginocchio SO$_8$
model (Sec.~\ref{so8}).  From the point of view of a link between
these models, an interesting comparison between the two analyses can
be made.

\section{Parameter symmetries of general Hamiltonians}
\label{general}

Following Refs.\cite{Shirokov1,Shirokov2}, we define the parameter
symmetry $\cal P$ of a given Hamiltonian $H(\lambda)$ depending on
a set of $m$ real parameters $\lambda\equiv\{\lambda_1,\lambda_2,
\dots\lambda_m\}$ as a mapping of the parameter space onto itself,
\begin{equation}
{\cal P}\,:\ \lambda\longmapsto\lambda'=f(\lambda)
\ ,\label{mapping}
\end{equation}
 such that $H(\lambda')$ is related to
$H(\lambda)$ by a similarity transformation,
\begin{equation}
H(\lambda')=U_{\lambda'\lambda}\ H(\lambda)\ U_{\lambda'\lambda}^{-1}
\ ,\label{similarity}
\end{equation}
where $U_{\lambda'\lambda}$ is a unitary operator. This is, of course,
equivalent with the requirement that Hamiltonians $H(\lambda)$ and
$H(\lambda')$ are isospectral: (i) from Eq.~(\ref{similarity})
it follows
that $U_{\lambda'\lambda}$ transforms the $H(\lambda)$ eigenvectors
$\ket{\psi_k(\lambda)}$ into the $H(\lambda')$ eigenvectors with the
same energy $E_k(\lambda)=E_k(\lambda')$, and (ii) an equality of
energies in $H(\lambda)=\sum_{k}\ket{\psi_k(\lambda)}E_k(\lambda)
\bra{\psi_k(\lambda)}$ and $H(\lambda')=\sum_{k}\ket{\psi_k(\lambda')}
E_k(\lambda')\bra{\psi_k(\lambda')}$ ensures that Eq.~(\ref{similarity})
is fulfilled with $U_{\lambda'\lambda}$ defined through $\ket{\psi_k
(\lambda')}=U_{\lambda'\lambda}\ket{\psi_k(\lambda)}$.

It is apparent that if ${\cal P}_1\equiv f_1(\lambda)$ and ${\cal P}_2
\equiv f_2(\lambda)$ are two parameter symmetries, then ${\cal P}_3
\equiv{\cal P}_1\circ{\cal
P}_2:\lambda\mapsto\lambda'=f_2(f_1(\lambda)) \equiv f_3(\lambda)$ is
also a parameter symmetry.  The parameter space of a system that
exhibits the parameter symmetry thus decomposes into subsets where
points of the same subset correspond to Hamiltonians with essentially
the same dynamics.  Each of these subsets, geometrically represented
by isolated points or a continuous manifold (for a smooth dependence
of the Hamiltonian on parameters), forms an equivalence class for
which any single member fully represents the whole class.  For
example, if the Hamiltonian $H(\lambda^0)$ is integrable, i.e., has
$f$ constants of motion $\{A_1,A_2,\dots A_f\}$ in involution with $f$
being the number of quantum degrees of freedom \cite{Zhang}, all the
Hamiltonians $H(\lambda)$ within the same equivalence class as
$H(\lambda^0)$ are integrable as well.  The integrals of motion are
simply given by $U_{\lambda\lambda^0}A_iU_{\lambda\lambda^0} ^{-1}$.
This feature of parameter symmetries is crucial for the study of
quantum chaos because it implies that perfectly regular dynamics can
be \lq\lq imported\rq\rq\ into parameter regions that might at first
be expected to be chaotic \cite{Kusnezov,Cejnar}.  There immediately
arise a plethora of questions related to the size and topological
structure of the equivalence classes in the parameter space.  The
answers, of course, depend on the particular Hamiltonian under study.

For a fixed pair $\lambda$ and $\lambda'$ relation (\ref{similarity})
implies a set of $n^2$ independent real equations -- $n$ being the
dimension of the Hilbert space -- to determine $n^2$ independent real
parameters of the unitary matrix $U_{\lambda'\lambda}$.  (Note that we
assume the Hamiltonian to be selfadjoint, but complex.)  In general,
the structure of the set of equations may very well produce no
solution.  To determine for a given $\lambda$ the range of $\lambda'$
for which a solution exists, it is convenient to consider the
diagonalized form of Eq.~(\ref{similarity}), i.e., the isospectral
condition.  This yields $n$ equations for $m$ variables $\lambda'$.
However, some of these equations might be identical.  This is
certainly the case if there exists some inherent degeneracy shared by
all Hamiltonians regardless of their parameter values.  Yet, not all
the mutually different eigenvalues of the Hamiltonian can be
considered independent (as elaborated in the next section).
Therefore, the number of relevant equations is given by the number
${\bar n}$ of {\em independent\/} energies.  If this is larger than
the number of parameters, ${\bar n}>m$, no parameter symmetries are
generally expected.  For $m={\bar n}$, typically single (discrete)
solutions $\cal P$ should be found.  Finally, a continuous variety of
similarity transformations may exist in an \lq\lq
overparameterized\rq\rq\ case, $m>{\bar n}$.

If $m\geq {\bar n}$, the expected dimensionality of manifolds
representing the equivalence classes of a given Hamiltonian is
$d=m-{\bar n}$. For instance, for a 2-dimensional (nondegenerate)
Hamiltonian dependent on 3 real parameters $\{\lambda_i\}_{i=1}^3$
the manifold containing the point $\lambda^0$ is formed by the
intersection of two surfaces $E_i(\lambda)-E_i(\lambda^0)=0,\ i=1,2$
in the 3-dimensional parameter space, which is a $d=1$ object,
a curve $C_{\lambda^0}$ crossing $\lambda^0$. Note that, the assumption
was made again that both the surfaces mentioned are continuous; we do
not trace here consequences of possible singularities in the
parametric dependence of $H(\lambda)$. In the case of a 4-parameter
Hamiltonian in 2 dimensions, the $\lambda^0$-containing manifold
will be a surface $S_{\lambda^0}$. For 2-parameter 2-dimensional
Hamiltonians, on the other hand, the $d=0$ equivalence classes in
the parameter space will typically consist of countable sets of
isolated points $P_{\lambda^0}=\{\lambda^0,\lambda^0{'},\lambda^0{''},
\dots\}$ (note that if the set is finite, it must be cyclic under
$\cal P$).

Assume now that there exists one or more mutually commuting and
{\em parameter-independent\/} constants of motion $\{A_1,\dots
A_q\}$ so that $[H(\lambda),A_i]=0,\ i=1\dots q$ {\em for all\/}
$\lambda$. It is then, of course, natural to require that an arbitrary
$\lambda\mapsto\lambda'$ transform of any of these integrals
(which is a new integral valid at $\lambda'$) must coincide with
the original. This implies that all similarity operators
$U_{\lambda'\lambda}$ in Eq.~(\ref{similarity}) are required to
commute with
all the $A_i$'s. Both the Hamiltonian and unitary matrices thus have
the same block-diagonal form, each block being associated with the
subspace of the total Hilbert space characterized by a particular
set of the $A_i$ quantum numbers. The above analysis can then be
applied either to the Hamiltonian as a whole or to each block
separately. Attention must be paid to the fact that some of the
submatrices may depend on a reduced number of parameters. As
a result, one can consider parameter symmetries in a particular
subset of blocks only. Moreover, the whole set of Hamiltonians
can possess an underlying symmetry represented by a group $G$.
In such a case, the unitary transformation should commute not only
with the Casimir operators, but also with all generators of $G$.
For the rotational symmetry, e.g., $U$ must commute with $J^2$,
$J_z$ and also with $J_+$ and $J_-$.

The implications of these general considerations depend very much on
the details of a particular situation, as is demonstrated in Sects.\
\ref{ibm} and \ref{so8} where we explore the
existence and nature of
possible parameter symmetries for two well-known nuclear models, the
interacting boson model (IBM) \cite{Iachello} and the SO$_8$ Ginocchio
model \cite{Ginocchio} (also in its fermion dynamical symmetry model
(FDSM) incarnation \cite{Wu}).

\section{Parameter symmetries of many-body Hamiltonians}
\label{manybody}

As was pointed out above, the general analysis of parameter
symmetries, based solely on the dimensionality of the problem, can
often be inconclusive or even misleading, because without knowledge of
the specific physics involved in the model it is hardly possible to
determine the number ${\bar n}$ of independent energies.
Turning to more specific examples of parameter symmetries in
bosonic or fermionic many-body systems, we now consider a
general Hamiltonian with one-, two-, three-, \dots $K$-body terms,
\begin{equation}
H(\Lambda)=\Lambda^{(0)}+\sum_{ij}\Lambda^{(1)}_{ij}a^{\dagger}_i
a_j+\sum_{ijkl}\Lambda^{(2)}_{ijkl}a^{\dagger}_ia^{\dagger}_ja_ka_l+
\sum_{ijklmn}\Lambda^{(3)}_{ijklmn}a^{\dagger}_ia^{\dagger}_j
a^{\dagger}_k a_la_m a_n +\dots+{\rm h.c.\ ,}
\label{Hmanybody}
\end{equation}
where $a^{\dagger}_i$ ($i=1,\dots s$) is the creation operator
of a particle in the $i$-th state (this state can also specify the
type of particle, like neutron or proton). Interaction strengths
$\Lambda\equiv\left\{\Lambda^{(0)},\{\Lambda^{(1)}_{ij}\},
\{\Lambda^{(2)}_{ijkl}\},\{\Lambda^{(3)}_{ijklmn}\},\dots\right\}$
form the general set of parameters of the $K$-body Hamiltonian
(they can be complex but the hermicity reduces the number of
independent real parameters -- for instance, there are not $2s^2$ but
only $s^2$ independent real one-body strengths $\Lambda^{(1)}$).

The Hamiltonian (\ref{Hmanybody}) conserves the total number of
particles, $N=\sum_ia^{\dagger}_ia_i$, as should any acceptable
similarity operator $U$. We thus have
\begin{equation}
Ua^{\dagger}_iU^{-1}=\sum_j\alpha^i_ja^{\dagger}_j
+\sum_{jkl}\beta^i_{jkl}a_ja_k^{\dagger}a_l^{\dagger}
+\sum_{jklmn}\gamma^i_{jklmn}a_ja_ka_l^{\dagger}a_m^{\dagger}
a_n^{\dagger}+\dots
\label{Ugen}
\end{equation}
where $\{\alpha^i_j\},\{\beta^i_{jkl}\},\{\gamma^i_{jklmn}\}\dots$
are some complex coefficients satisfying constraints imposed by the
unitarity. Eq.~(\ref{Ugen}) determines the most general particle-number
conserving similarity transformation in the many-body Fock space.
However, it does not constitute a parameter symmetry of a $K$-body
Hamiltonian if higher-order terms with coefficients $\beta^i_{jkl},
\gamma^i_{jklmn}\dots$ are included. This is so because the higher-order
terms increase the maximum order, $K$, of interactions in the Hamiltonian.
Parameter symmetries of the Hamiltonian (\ref{Hmanybody}) with $K$ finite
can thus be specified by the simplified version
\begin{equation}
Ua^{\dagger}_iU^{-1}=\sum_j\alpha^i_ja^{\dagger}_j
\label{U1}
\end{equation}
of Eq.~(\ref{Ugen}),
with the unitarity constraint $\sum_k\alpha^i_k\alpha^{j*}_k=\delta_{ij}$.
Namely, Eq.~(\ref{U1}) clearly yields the following $\Lambda\mapsto
\Lambda'$ mapping:
\begin{eqnarray}
\Lambda^{(0)}{'} & = & \Lambda^{(0)}\nonumber\\
\Lambda^{(1)}{'}\!\!\!\!\!\!_{ij}\ & = & \sum_{kl}\alpha^k_i\alpha^{l*}_j
\Lambda^{(1)}_{kl}\ ,\nonumber\\
\Lambda^{(2)}{'}\!\!\!\!\!\!\!_{ijkl}\, & = & \sum_{mnpq}\alpha^m_i\alpha^n_j
\alpha^{p*}_k\alpha^{q*}_l\Lambda^{(2)}_{mnpq}\ ,\label{trans1}\\
\Lambda^{(3)}{'}\!\!\!\!\!\!\!_{ijklmn} & = & \dots \nonumber
\end{eqnarray}

If there are some additional global integrals of motion besides $N$,
the right-hand side of Eq.~(\ref{U1}) can only mix the creation
operators $a_j^{\dagger}$ that carry the same values of these
integrals as $a_i^{\dagger}$ on the left-hand side, i.e.,
$\alpha^i_j=0$ if $i$ and $j$ label states that differ in one or more
conserving quantum numbers.  This most obviously applies to the
angular momentum $J^2$ and its projection $J_z$ in the case of
underlying rotational symmetry.  Yet other conserved quantities (like
charge etc.) may be relevant.  Moreover, as all generators of the
symmetry group $G$ must commute with $U$, some additional constraints
emerge.  In the case of rotational symmetry one gets relations of the
type $\alpha^i_j=\alpha^{i_{\pm}} _{j_{\pm}}$, where $i_{\pm}$ and
$j_{\pm}$ represent states obtained by applying $J_+$ or $J_-$ to $i$
and $j$, respectively.

As follows from the previous discussion, the particle-number
conserving many-body Hamiltonian in its most general parameterization
(\ref{Hmanybody}) always exhibits parameter symmetries (\ref{trans1}).
Even if the form of Eq.~(\ref{Hmanybody}) is further restricted by
imposing additional integrals of motion and symmetries, the similarity
transformation can be constructed whenever Eq.~(\ref{U1}) allows,
while respecting all the above-discussed constraints, the construction
of a nontrivial transformation of the single-particle operators.  This
is so in spite of the fact that the number of independent real
parameters composed from the $\Lambda$'s may be (and usually is)
smaller than the number of mutually different many-body energies.
Obviously, not all of these energies are necessarily independent, as
illustrated by the simplest example of a Hamiltonian with just
single-particle interactions for which all many-particle energies
appear as simple combinations of the set of single-particle energies.
In fact, the number of independent energies of a given many-body
Hamiltonian must -- by definition -- be smaller than the number of
parameters in the {\em most general\/} parameterization.  If the
$\Lambda$'s in Eq.~(\ref{Hmanybody}) are made dependent on a smaller
set of parameters, $\{\lambda\}$, the quest for parameter symmetries
in the reduced parameter space translates into the search of those
transformations $\lambda\mapsto\lambda'$ that accommodate the mapping
in Eq.~(\ref{trans1}).

It should be pointed out that besides the standard single-particle
transformations in Eq.~(\ref{U1}), a formal exchange of creation and
annihilation operators (particles and holes) in the Hamiltonian was
also considered in Refs.\cite{Shirokov1,Shirokov2} as a possible
transformation leading to parameter symmetries. Note that the
inclusion of such inverted terms on the right-hand side of
Eq.~(\ref{Ugen}) would preserve the particle-number conservation
of the transformed Hamiltonian. However, the transformation itself
with these additional terms is clearly nonunitary (the basis of the
$N=1$ subspace is mapped onto states that all have a nonzero overlap
with the vacuum and thus are not orthonormal). That is why we do not
include such transformations into our analysis, although they may be
relevant if only $N\geq 2$ subspaces are considered.

\section{The interacting boson model-1}
\label{ibm}

As a simple and well-studied example \cite{Shirokov1}, let us consider
first the IBM-1 \cite{Iachello}. It is formulated
in terms of two kinds of bosons, $s$ and $d$, with angular momenta
0 and 2, respectively, that interact via a Hamiltonian of the type
(\ref{Hmanybody}) with only one- and two-body terms. In addition, the
Hamiltonian is assumed to be invariant under rotations and the time
reversal. Its general form is given by the following expression,
\begin{eqnarray}
H(\lambda)=k_0+k_1C_1({\rm U}_5)+k_2C_2({\rm U}_5)+k_3C_2({\rm SO}_5)
 & + & k_4C_2({\rm SO}_3)\nonumber\\
& + & k_5C_2({\rm SO}_6)+k_6C_2({\rm SU}_3)\ ,
\label{Hibm}
\end{eqnarray}
where $\lambda\equiv\{k_0,\dots k_6\}$ are real parameters, weights
of Casimir operators corresponding to groups involved in chains connecting
the dynamical group U$_6$ with the symmetry group SO$_3$ (their explicit
form can be found, e.g., in Ref.\cite{Shirokov2}).

The dimensional analysis would indicate that parameter symmetries
can hardly be found in subspaces with large total boson number.
Indeed, the number of different energies (there is always
the degeneracy associated with SO$_3$ symmetry) exceeds the number of
parameters for $N\geq 3$. However, the above analysis related
specifically to many-body Hamiltonians shows that the parameter
symmetry exists in all $N$-subspaces. Let us consider the boson
transformations according to Eq.~(\ref{U1}). Clearly, because $J^2$
and $J_z$ are parameter-independent constants of motion, we can only
consider transformations $Us^{\dagger}U^{-1}=e^{i\phi_s}s^{\dagger}$
and $Ud_{\mu}^{\dagger}U^{-1}=e^{i\phi_{\mu}}d_{\mu}^{\dagger}$,
where $s^{\dagger}$ and $d_{\mu}^{\dagger}$ create, respectively,
an $s$-boson and a $d$-boson with $J_z$ projection $\mu=-2\dots+2$,
while $\phi_s$ and $\{\phi_{\mu}\}$ represent arbitrary real phases. 
However, the fact that $U$ must commute also with the remaining 
SO$_3$ generators, $J_+$ and $J_-$, results in the requirement that 
the $d$-boson phases are independent of $\mu$, i.e., $\phi_{\mu}=
\phi_d$. Furthermore, since the Hamiltonian (\ref{Hibm}) is invariant
under a global gauge transformation of all creation and annihilation
operators, the only remaining parameter is the relative phase between
$s$- and $d$-bosons, $\Delta\phi=\phi_s-\phi_d$.  Without any
loss of generality we can set $\phi_d=0$, thus
\begin{equation}
Us^{\dagger}U^{-1}=e^{i\phi_s}s^{\dagger}\qquad,\qquad
Ud_{\mu}^{\dagger}U^{-1}=d_{\mu}^{\dagger}
\label{U1ibm}
\end{equation}
(or, equivalently, $\phi_s=0$ and $\phi_d \ne 0$). The reality of
coefficients in the
Hamiltonian (the time-reversal invariance) allows only some discrete
values of $\phi_s$ \cite{Iachello,Isacker,Shirokov1,Shirokov2}, namely
$\phi_s=0, \pi$ for $k_6\neq 0$ and $\phi_s=0, \pm\pi/2, \pi$ for
$k_6=0$.

From our general analysis, we therefore arrive at a discrete set of
similarity transformations
of the Hamiltonian (\ref{Hibm}) that exactly coincide with the gauge
transformations described in earlier work \cite{Iachello,Isacker}.
After some algebra with the Casimir operators in Eq.~(\ref{Hibm})
\cite{Kusnezov,Cejnar,Shirokov1,Shirokov2}, one derives the following
mapping corresponding to Eq.~(\ref{U1ibm}) with the above-given discrete
values of $\phi_s$:
\begin{equation}
\biggl(k'_0,k'_1,k'_2,k'_3,k'_4,k'_5,k'_6\biggr)=\left\{
\begin{array}{r}
\biggl(k_0,\ k_1\!+\!2k_6,\ k_2\!+\!2k_6,\ k_3\!-\!6k_6,\ k_4\!+
\!2k_6,\ k_5\!+\!4k_6,\ -k_6\biggr) \\
\text{\quad if $k_6\neq 0$},\\
\biggl(k_0\!+\!10Nk_5,\ k_1\!+\!4(N\!+\!2)k_5,\ k_2\!-\!4k_5,\ k_3\!+
\!2k_5,\ k_4,\ -k_5,\ 0\biggr) \\
\text{\quad if $k_6=0$}.
\end{array}
\right.
\label{Shiroko}
\end{equation}
This is the parameter symmetry given by Shirokov {\it et al.\/}
\cite{Shirokov1}. It implies, in particular, that any Hamiltonian
$H(\lambda)$ possessing the SU$_3$ dynamical symmetry, $\lambda=
(k_0,0,0,0,k_4,0,k_6)$, has an isospectral partner $H(\lambda')$ 
with $\lambda'=(k_0,2k_6,2k_6,-6k_6,k_4\!+\!2k_6,4k_6,-k_6)$, which
is therefore integrable in spite of nonzero admixtures of all the 
U$_5$, SO$_6$ and SU$_3$ dynamical symmetries [the Hamiltonian
$H(\lambda')$ is said to have the so called ${\overline{\rm SU}_3}$ 
or ${\rm SU}_3^*$ \lq\lq hidden\rq\rq\ dynamical symmetry]. 
Similarly, the SO$_6$ Hamiltonians with $\lambda=(k_0,0,0,k_3,
k_4,k_5,0)$ have the ${\overline{\rm SO}_6}$ (or ${\rm SO}_6^*$) 
isospectral partners at $\lambda'=(k_0\!+\!10Nk_5,4(N\!+\!2)k_5,
-4k_5,k_3\!+\!2k_5,k_4,-k_5,0)$, i.e., in the U$_5$--SO$_6$ 
transitional region.

In fact, Eq.~(\ref{Shiroko}) represents a single mapping of
the parameter space onto itself, a mapping discontinuous at
$k_6=0$. However, since the origin of this discontinuity
is the extension of the allowed $\phi_s$ values at $k_6=0$
(see above), Eq.~(\ref{Shiroko}) is more appropriately viewed as 
two separate continuous mappings, the first valid in the whole
parameter space and yielding just the identity for $k_6=0$,
the second applicable only in the $k_6=0$ subspace.
The fact that the subset of Hamiltonians with no admixture of the
SU$_3$ Casimir operator is invariant under the transformation
(\ref{Shiroko}) is important as all these Hamiltonians are known 
to be integrable \cite{Whelan} and this property is thus not 
propagated into other regions of the parameter space. 

Under the transformation (\ref{Shiroko}), the full 7-dimensional 
IBM-1 parameter space decomposes into pairs of points that 
constitute the dynamical equivalence classes of the model. Two 
consecutive transformations (\ref{Shiroko}) form the identity. 
Of course, less complex parameterizations (such as the one in
Refs.\cite{Cejnar,Whelan}) typically contain at most one of the
two isospectral Hamiltonians present in the complete parameter
space. Let us stress that the equivalence of Hamiltonians
connected by Eq.~(\ref{Shiroko}) does not imply the same
transition rates if a fixed, parameter-independent set of 
transition operators is used. However, the transition rates
in both points $\lambda$ and $\lambda'$ will apparently be 
equal if the transition operators at $\lambda'$ are chosen 
to be the $U_{\lambda'\lambda}$-transforms of the transition 
operators at $\lambda$. A detailed discussion of this point 
(making a link with the so called consistent-$Q$ formalism) 
can be found in Ref.\cite{Shirokov1}.

Our analysis leads us to disagree with the statement made in
Ref.\cite{Shirokov1}
concerning an additional IBM-1 parameter symmetry that does
not result from a transformation of the type (\ref{U1}). This
parameter symmetry is allegedly constructed in three steps: (i)
the expansion of the Hamiltonian (\ref{Hibm}) in terms of the
set of Casimir operators where $C_2({\rm SO}_6)$ is replaced by
$C_2({\overline{\rm SO}_6)}$ (the group ${\overline{\rm SO}_6}$
differs from the \lq\lq standard\rq\rq\ SO$_6$ by the above gauge
transformation with $\phi_s=\pm\pi/2$), (ii) the application of
the transformation (\ref{Shiroko}) to the expansion obtained in
the first step, and (iii) the reverse decomposition of
$C_2({\overline{\rm SO}_6)}$ in the resulting expression into
standard Casimir operators. Indeed, when literally following these
steps, one finds a Hamiltonian that differs from the one obtained
by merely applying Eq.~(\ref{Shiroko}). However, this Hamiltonian
is not isospectral with the original one because the transformation
in Eq.~(\ref{Shiroko}) does {\em not\/} represent a parameter
symmetry for the Hamiltonian decomposition in terms of
$C_2({\overline{\rm SO}_6)}$. If it were so, one could repeatedly
apply the new transformation and the one from Eq.~(\ref{Shiroko})
yielding a chain of new equivalent Hamiltonians in the parameter
space. The dynamical equivalence classes would then be infinite
(although
countable) sets. However, this is not the case and the mapping
(\ref{Shiroko}) represents {\em the only\/} parameter symmetry
of the IBM-1.

\section{The Ginocchio SO$_8$ model}
\label{so8}

To explore the microscopic origin of the interacting boson model and
its success as a phenomenological model, one has to link $s-$ and $d-$
bosons
to nucleon pairs. A promising perspective was offered by the
Ginocchio SO$_8$ model \cite{Ginocchio}, later generalized to the
fermion dynamical symmetry model \cite{Wu}, and formulated in terms
of $s$- and $d$-bosons by Geyer and Hahne \cite{GH81}. In the
Ginocchio
model, an even number ($2N$) of fermions is considered in a shell of
single-particle states with total angular momenta $j$ decomposed as
${\vec j}=\vec{k}+{\vec\frac{3}{2}}$, where $\vec k$ is the so-called
pseudo-orbital angular momentum ($k$ is a positive integer) and
${\vec\frac{3}{2}}$ is termed the pseudospin. Under this restriction,
the total
angular momentum ${\vec J}={\vec j_1}+{\vec j_2}$ of a nucleon pair
can only be $J=0$ or 2 if ${\vec k}_1+{\vec k}_2$ couples to zero
in each pair. These $S$- and $D$-fermion pairs are counterparts of
the IBM $s$- and $d$-bosons (see Ref.\ \cite{GH81}) .

The Ginocchio fermionic Hamiltonian involves the usual one- plus
two-body interaction terms.  The general form that conserves the total
angular momentum is,
\begin{equation}
H(\Lambda)=\Lambda^{(0)}+\sum_{j}\Lambda^{(1)}_j\left(\sum_m
(-)^{m-j}a_{jm}^{\dagger}{\tilde a}_{j-m}\right)+
\sum_{j_1j_2j_3j_4J}\Lambda^{(2)}_{j_1j_2j_3j_4J}\left(\sum_M
(-)^{M-j_3-j_4}[a^{\dagger}_{j_1}a^{\dagger}_{j_2}]^J_M
[{\tilde a}_{j_3}{\tilde a}_{j_4}]^J_{-M}\right)\ ,
\label{Hferm}
\end{equation}
with $[A_{j_1}B_{j_2}]^J_M=\sum_{m_1,m_2}(j_1m_1j_2m_2|JM)A_{j_1m_1}
B_{j_2m_2}$. Here the single particle operators
$a^{\dagger}_{jm}$ and ${\tilde a}_{jm}=(-)^{j+m}a_{j-m}$ are
restricted to
$j=|k-\frac{3}{2}|,\dots(k+\frac{3}{2})$ and $m=-j,\dots+j$.
Hermicity of the Hamiltonian requires $\Lambda^{(0)}$ and
$\{\Lambda^{(1)}\}$ to be real, while the
interaction strengths $\{\Lambda^{(2)}\}$ satisfy the
condition $\Lambda^{(2)}_{j_1j_2j_3j_4J}=\Lambda^{(2)*}_{j_4j_3j_2j_1J}$.
We also set $\Lambda^{(2)}_{j_1j_2j_3j_4J}=\Lambda^{(2)}_{j_2j_1j_4j_3J}$,
as naturally follows from symmetry properties of the two-body operators
in Eq.~(\ref{Hferm}).

Before discussing parameter symmetries of the more specific Ginocchio
Hamiltonian, let us consider the ones of the most general Hamiltonian
(\ref{Hferm}). From the previous sections we know that the relevant
transformations must be of the following form,
\begin{equation}
Ua^{\dagger}_{jm}U^{-1}=e^{i\phi_j}a^{\dagger}_{jm}\ ,
\label{Uferm}
\end{equation}
where $\phi_j$ are arbitrary real phases. This clearly leads to
\begin{eqnarray}
\Lambda^{(0)}{'} & = & \Lambda^{(0)}\ ,\label{transferm0}\\
\Lambda^{(1)}_j{'} & = & \Lambda^{(1)}_j\ ,\label{transferm1}\\
\Lambda^{(2)}{'}\!\!\!\!\!\!\!\!_{j_1j_2j_3j_4J} & = &
e^{i(\phi_{j_1}+\phi_{j_2}-\phi_{j_3}-\phi_{j_4})}
\Lambda^{(2)}_{j_1j_2j_3j_4J}\ .\label{transferm2}
\end{eqnarray}
Suppose now that the Hamiltonian (\ref{Hferm}) is invariant under
the time reversal. The time reversal operator $T$ is antiunitary
and we choose the convention with $Ta^{\dagger}_{jm}T^{-1}=(-)^{j+m}
a^{\dagger}_{j-m}$, $T{\tilde a}_{jm}T^{-1}=(-)^{j+m}{\tilde a}_{j-m}$.
In addition to the hermicity constraints, we then arrive at the
further constraint that
the coefficients $\Lambda^{(2)}_{j_1j_2j_3j_4J}$
are either real (if $j_1+j_2+j_3+j_4$ is even) or imaginary (if
$j_1+j_2+j_3+j_4$ is odd).

These results lead to a severe restriction of possible values of phases
in Eq.~(\ref{Uferm}). Namely, from Eq.~(\ref{transferm2}) we see that
the conservation of purely real or imaginary character of the
two-body strengths requires
$\phi_{j_1}+\phi_{j_2}-\phi_{j_3}-\phi_{j_4}=
n_{j_1j_2j_3j_4}\pi$ with $n_{j_1j_2j_3j_4}=0,\pm 1,\pm 2,\dots$ for
each $j_1,j_2,j_3,j_4$. This will certainly be so if individual phases
$\phi_j$ differ by multiples of $\pi$. As the global gauge is
irrelevant and as only phase values modulo $2\pi$ suffice, we end up
with transformations generated by various permutations of phases 0
and $\pi$. For instance, if $k\geq 2$, the four phases $\{\phi_{k-
\frac{3}{2}},\dots\phi_{k+\frac{3}{2}}\}\equiv\{\phi_1,\dots\phi_4\}$
can take any combination of values from the following set:
\begin{equation}
(\phi_1,\phi_2,\phi_3,\phi_4)=(0,0,0,\pi),(0,0,\pi,0),(0,\pi,0,0),
(\pi,0,0,0),(0,0,\pi,\pi),(0,\pi,0,\pi),(\pi,0,0,\pi).
\label{permut}
\end{equation}
Note that the remaining combinations are just $0\rightleftharpoons\pi$
conjugates of the ones given above and produce therefore equivalent
transformations. Each of the 7 possibilities in Eq.~(\ref{permut})
generates a specific parameter symmetry that operates in the entire
parameter
space of the most general Hamiltonian (\ref{Hferm}). It should be noted,
however, that for Hamiltonians with $\Lambda^{(2)}_{j_1j_2j_3j_4J}=0$ for
some particular combinations of angular momenta (i.e., in some parameter
subspaces), additional parameter symmetries can be possible. Let us
recall that a similar situation was met in the IBM for $k_6=0$, which
in the present language corresponds to $\Lambda^{(2)}_{22202}=0$.

The Ginocchio Hamiltonian is not as general as the one in
Eq.~(\ref{Hferm}). It turns out \cite{Ginocchio} that the $S$ and $D$
fermionic pair operators belong to the SO$_8$ algebra. The model
Hamiltonian is thus built exclusively from generators of this algebra,
i.e., possesses the SO$_8$ dynamical symmetry. The pair
creation operators are defined in the following way,
\begin{eqnarray}
S^{\dagger} & = & \frac{1}{\sqrt{2\Omega}}\sum_j\sqrt{2j+1}
\ [a_j^{\dagger}a_j^{\dagger}]^0_0\ ,
\label{Sdef}\\
D^{\dagger}_M & = & \frac{1}{\sqrt{\Omega}}\sum_{j_1,j_2}
(-)^{j_1+k+\frac{3}{2}}\sqrt{(2j_1+1)(2j_2+1)}
\left\{\begin{array}{ccc}
j_1 & j_2 & 2 \\
\frac{3}{2} & \frac{3}{2} & k
\end{array}\right\}
[a_{j_1}^{\dagger}a_{j_2}^{\dagger}]^2_M\ ,
\label{Ddef}
\end{eqnarray}
where $\Omega$ is the maximum number of nucleon pairs in a fully occupied
shell, $2\Omega=\sum_j(2j+1)=4(2k+1)$. The corresponding pair
annihilation
operators are Hermitian conjugates of Eqs.~(\ref{Sdef}) and
(\ref{Ddef}).
The remaining SO$_8$ generators are four multipole operators
\begin{equation}
P^r_M=2\sum_{j_1,j_2}(-)^{r+j_1+k+\frac{3}{2}}\sqrt{(2j_1+1)(2j_2+1)}
\left\{\begin{array}{ccc}
j_1 & j_2 & r \\
\frac{3}{2} & \frac{3}{2} & k
\end{array}\right\}
[a_{j_1}^{\dagger}{\tilde a}_{j_2}]^r_M
\qquad (r=0,1,2,3).
\label{Pdef}
\end{equation}
The Ginocchio SO$_8$ Hamiltonian is expressed in terms of
the definitions~(\ref{Sdef})--(\ref{Pdef}),
\begin{equation}
H(\lambda)=E_0+G_0\,S^{\dagger}S+G_2\sum_M D^{\dagger}_MD_M+\frac{1}{4}
\sum_{r=1}^3b_r\sum_M(-)^MP^r_MP^r_{-M}\ ,
\label{HGino}
\end{equation}
where $\lambda\equiv\{E_0,G_0,G_2,b_1,b_2,b_3\}$ are real control
parameters. Expressed in the form of Eq.~(\ref{Hferm}), the
Hamiltonian (\ref{HGino}) yields the following strength coefficients:
\begin{eqnarray}
& & \Lambda^{(0)}=E_0
\label{Lam0}
\\
& & \Lambda^{(1)}_j=\sum_{r=1}^3(2r+1)\sum_{j'}(2j'+1)
\left\{\begin{array}{ccc}
j & j' & r \\
\frac{3}{2} & \frac{3}{2} & k
\end{array}\right\}^2
\,b_r
\label{Lam1}
\\
& & \Lambda^{(2)}_{j_1j_2j_3j_4J}=
\delta_{J0}\,\delta_{j_1j_2}\delta_{j_3j_4}\,\frac{1}{2\Omega}
\sqrt{(2j_1+1)(2j_3+1)}\ \,G_0+
\nonumber
\\
& & \qquad\quad\delta_{J2}(-)^{j_1+j_4+1}\frac{1}{\Omega}
\sqrt{(2j_1+1)(2j_2+1)(2j_3+1)(2j_4+1)}
\left\{\begin{array}{ccc}
j_1 & j_2 & 2 \\
\frac{3}{2} & \frac{3}{2} & k
\end{array}\right\}
\left\{\begin{array}{ccc}
j_3 & j_4 & 2 \\
\frac{3}{2} & \frac{3}{2} & k
\end{array}\right\}
\ G_2+
\label{Lam2}
\\
& & \qquad (-)^{j_1+j_4}\sqrt{(2j_1+1)(2j_2+1)(2j_3+1)(2j_4+1)}
\,\sum_{r=1}^3(2r+1)
\left\{\begin{array}{ccc}
j_1 & j_3 & r \\
\frac{3}{2} & \frac{3}{2} & k
\end{array}\right\}
\left\{\begin{array}{ccc}
j_2 & j_4 & r \\
\frac{3}{2} & \frac{3}{2} & k
\end{array}\right\}
\left\{\begin{array}{ccc}
j_1 & j_2 & J \\
j_4 & j_3 & r
\end{array}\right\}
\ b_r
\nonumber
\end{eqnarray}
Note that the one-body terms (\ref{Lam1}) result from the normal
ordering of the last term in Eq.~(\ref{HGino}). It is also clear
that the assumption concerning $S$- and $D$-pairs does not restrict
the two-body matrix elements (\ref{Lam2}) to $J=0,2$ only.
Apparently, all the two-body terms (\ref{Lam2}) fulfill the
hermicity condition and, in addition, are real. Indeed, because the
$j_1\rightleftharpoons j_2, j_3\rightleftharpoons j_4$ symmetry
implies that $(-)^{j_1+j_4}=(-)^{j_2+j_3}$ [Eqs.~(\ref{Ddef}) and
(\ref{Pdef}) are invariant under $j_1\rightleftharpoons j_2$], the
right-hand side of Eq.~(\ref{Lam2}) is nonzero only for even values
of the sum $j_1+j_2+j_3+j_4$.

It is now simple to see that no parameter mapping
$\lambda\mapsto\lambda'$
can realize the gauge transformation (\ref{transferm0})--(\ref{transferm2}).
Firstly, Eqs.~(\ref{transferm0}) and (\ref{Lam0}) yield $E_0'=E_0$,
while
$b_r'=b_r\ (r=1,2,3)$ follows from Eqs.~(\ref{transferm1}) and (\ref{Lam1})
as coefficients at $b_r$ in (\ref{Lam1}) are positive. Since
$\Lambda^{(2)}{'}\!\!\!\!\!\!\!\!_{j_1j_1j_2j_20}=\Lambda^{(2)}
_{j_1j_1j_2j_20}$ also follows from Eqs.~(\ref{transferm2}) and
(\ref{permut}), we
furthermore find $G_0'=G_0$. The only remaining parameter, $G_2$, can
clearly
not fulfill the consistent transformation (\ref{transferm2}) of
{\it all}
two-body strengths (for instance, it can only affect the terms with $J=2$).

These results are interesting from the viewpoint of the known 
correspondence between the SO$_8$ model and the IBM-1 \cite{GH81,Kim}.
Given that these models are assumed to describe basically the same 
physics, one can ask why the SO$_8$ parameter space fails to 
accommodate isospectral Hamiltonians in contrast to the IBM-1 space. 
To answer this question one has to specify the method used to link 
both models. Among various fermion-boson mapping techniques 
\cite{Klein}, the Dyson mapping is favored by the fact that it 
transforms the two-body fermionic Hamiltonian (\ref{HGino}) into 
a two-body bosonic Hamiltonian. However, the subsequent hermitization 
of the bosonic Hamiltonian \cite{Kim,Geyer} is necessary, which 
seems to be possible -- without introducing three- and more-body 
boson interactions -- only for a certain subset of the SO$_8$ 
parameter space \cite{Kim}. It means that the dynamical equivalence 
of the SO$_8$ model and the IBM-1 in terms of the link 
$(E_0,G_0,G_2,b_1,b_2,b_3)\mapsto(k_0,\dots k_6)$ between parameters 
in Eqs.~(\ref{HGino}) and (\ref{Hibm}), respectively, can probably 
be established for this limited SO$_8$ parameter subset only. Let
us note that some uncertainty in the last statement results from 
the fact that there is, strictly speaking, no proof that another 
hermitization procedure (also preserving the two-body character 
of interactions as the one from Ref.\cite{Kim}) cannot be more
successful in the problematic parameter region.

On the other hand, by inspection of the mapping formulas in 
Refs.\cite{Kim,Geyer} it becomes apparent that not each Hamiltonian 
(\ref{Hibm}) can be mapped from a Hamiltonian of the form 
(\ref{HGino}). In this sense, the SO$_8$ parameter space is 
smaller than the IBM-1 space, i.e., the SO$_8\to$\ IBM-1 parameter 
mapping is not surjective (onto) but only injective (into the 
IBM-1 space). The absence of parameter symmetries in the 
Ginocchio model then indicates that the image of the SO$_8$ 
parameter space in the IBM-1 space contains no equivalence 
classes, or in other words, that out of each pair of the equivalent 
IBM-1 Hamiltonians {\em at most one\/} has a counterpart 
within the SO$_8$ space. Note, however, that the bosonic gauge 
transformations (\ref{U1ibm}) can be easily realized by choosing 
an appropriate phase convention in the Dyson mapping.

\section{Conclusions}

We have discussed parameter symmetries of general quantum
many-body systems. Identifying such symmetries on the basis of
constraints which result from a comparison of the number of
parameters
with the number of independent eigenvalues of the Hamiltonian, is not
practical because of the difficulty to determine the latter in
general. It
was shown, however, that the restrictions imposed upon the similarity
transformations $U$, namely the commutation of $U$ with (a) all
parameter-independent integrals of motion and (b) all generators of
the symmetry group, are sometimes sufficient for this determination.

From the above considerations we identified and proposed for many-body
Hamiltonians conserving the total number of particles
the following procedure: (i) consider single-particle
transformations of the type (\ref{U1}) conserving all the model integrals
of motion; (ii) apply the commutation rules under (b) to further
restrict these transformations; (iii)  exclude global gauge
transformations that only lead to the trivial mapping $\lambda\mapsto
\lambda$ (such transformations belong to the Abelian symmetry group).

This procedure applied to the interacting boson model-1 showed that
the parameter symmetry (\ref{Shiroko}), derived in
Ref.\cite{Shirokov1}, is the only parameter symmetry of this model.
In fact, since the Hamiltonian (\ref{Hibm}) is the most general
rotationally invariant one- plus two-body Hamiltonian with $s$- and
$d$-boson degrees of freedom, our general analysis already indicates
that parameter symmetries should be a natural ingredient of the model.
In contrast, a similar analysis of the Ginocchio SO$_8$ model
disclosed that the parameterization (\ref{HGino}) is too restrictive
to allow for any parameter symmetries, although such symmetries exist
in the more general parameterization (\ref{Hferm}).  These results of
course do not contradict any aspect of the relationship between the
SO$_8$ model and an IBM-like $s$- and $d$-boson counterpart based on
boson-fermion mappings, where mapped Hamiltonians generally represent
a restricted subset of the most general form (\ref{Hibm}).

Let us stress finally that the analysis would become much more
complicated if we were to consider many-body Hamiltonians with
interactions of arbitrary order. Eq.~(\ref{Ugen}) would then have to
be applied in its general form and no obvious insight seems available
to do so.

\acknowledgements{This work was supported by the S.A. National Research 
Foundation under grants GUN 2047181 and GUN 2044653 and partly by the
Grant Agency of Czech Republic under grant 202/99/1718.}

\end{document}